\documentclass[a4paper, 11pt]{article} % Font size (can be 10pt, 11pt or 12pt) and paper size (remove a4paper for US letter paper)

\usepackage[protrusion=true,expansion=true]{microtype} % Better typography
\usepackage{graphicx} % Required for including pictures
\usepackage{wrapfig} % Allows in-line images

\usepackage{mathpazo} % Use the Palatino font
\usepackage[T1]{fontenc} % Required for accented characters

\usepackage{rotating}
\usepackage{url}
\usepackage{hyperref}
\usepackage{float}

\newcommand\eemph[1]{\textbf{#1}}

\makeatletter
\renewcommand\@biblabel[1]{\textbf{#1.}} % Change the square brackets for each bibliography item from '[1]' to '1.'
\renewcommand{\@listI}{\itemsep=0pt} % Reduce the space between items in the itemize and enumerate environments and the bibliography

\renewcommand{\maketitle}{ % Customize the title - do not edit title and author name here, see the TITLE block below
\begin{flushright} % Right align
{\LARGE\@title} % Increase the font size of the title

\vspace{50pt} % Some vertical space between the title and author name

{\large\@author} % Author name
\vspace{40pt} % Some vertical space between the author block and abstract
\end{flushright}
}

\title{\textbf{Metamorphic Domain-Specific Languages}\\ % Title
A Journey Into the Shapes of a Language} % Subtitle

\author{\textsc{Mathieu Acher, Benoit Combemale} % Author
\\
{\textit{University of Rennes 1, Inria, IRISA}
\\
\textit{France}
\\
\textsf{\{mathieu.acher, benoit.combemale\}@irisa.fr}
\\
\textsc{\\Philippe Collet}
\\{\textit{University of Nice Sophia Antipolis, I3S laboratory / CNRS}}
\\
\textit{France}
\\
\textsf{philippe.collet@unice.fr}
}} % Institution

\begin{document}

\maketitle % Print the title section

%----------------------------------------------------------------------------------------
%	ABSTRACT AND KEYWORDS
%----------------------------------------------------------------------------------------

%\renewcommand{\abstractname}{Summary} % Uncomment to change the name of the abstract to something else

\begin{abstract}
External or internal domain-specific languages (DSLs) or (fluent) APIs?  Whoever you are -- a developer or a user of a DSL -- you usually have to choose your side; you should not! What about metamorphic DSLs that change their shape according to your needs?

We report on our 4-years journey of providing the "right" support (in the domain of feature modeling), leading us to develop an external DSL, different shapes of an internal API, and maintain all these languages. A key insight is that there is no one-size-fits-all solution or no clear superiority of a solution compared to another. On the contrary, we found that it does make sense to continue the maintenance of an external and internal DSL. 

The vision that we foresee for the future of software languages is their ability to be self-adaptable to the most appropriate shape (including the corresponding integrated development environment) according to a particular usage or task. We call metamorphic DSL such a language, able to change from one shape to another shape.
\end{abstract}

\hspace*{3,6mm}\textit{Keywords:} programming, domain-specific languages, metamorphic % Keywords

\vspace{30pt} % Some vertical space between the abstract and first section

%----------------------------------------------------------------------------------------
%	ESSAY BODY
%----------------------------------------------------------------------------------------

\section*{Introduction}

Domain-specific languages (DSLs) are more and more used to leverage business or technical domain expertise inside complex software-intensive systems. DSLs are usually simple and little languages, focused on a particular problem or aspect of a (software) system. Outstanding examples of DSLs are plentiful: Makefiles for building software, Matlab for numeric computations, Graphviz language for drawing graphs, or SQL (Structured Query Language) for databases.

People find DSLs valuable because a well-designed DSL can be much easier to program with than a traditional library. The case of SQL is a typical example. Before SQL was conceived, querying and updating relational databases with available programming languages leads to a huge semantic gap between data and control processing. With SQL, users can write a query in terms of an implicit algebra without knowing the internal layout of a database. Users can also benefit from performance optimization: a query optimizer can determine the most efficient way to execute a given query. 
Another benefit of DSLs is their capacity at improving communication with domain experts~\cite{mernik2005,FowlerDSLBook,voelter2013}, which is an important tool for tackling one of the hardest problems in software development. 

DSLs are also plain languages, in the sense that many difficult design decisions must be taken during their construction and maintenance~\cite{mernik2005}, and that they can take different shapes: plain-old to more fluent Application Programming Interface (APIs) ; internal or embedded DSLs written inside an existing host language ; external DSLs with their own syntax and domain-specific tooling. 
To keep it simple, a useful and common distinction is to consider that a DSL comes in two main shapes~\cite{FowlerDSLBook}: external and internal. When an API is primarily designed to be readable and to "flow", we consider it is also a DSL.

\begin{figure*}
\center
\includegraphics[scale=0.55]{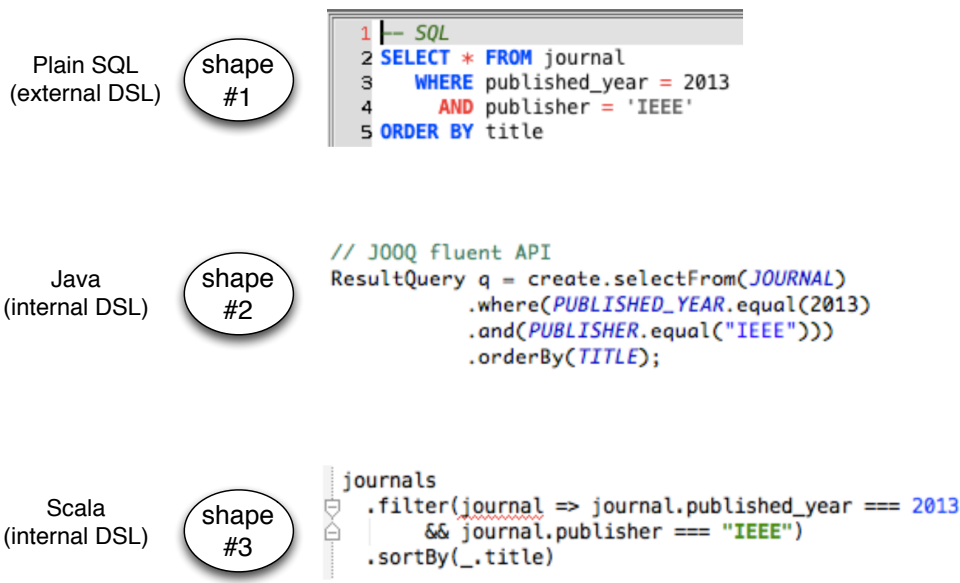}
\caption{Three SQL \emph{shapes}: plain SQL, JOOQ fluent API in Java, Slick API in Scala\label{fig:sqlshapes}}
\end{figure*}

It is interesting to note that SQL -- invented in 1974 and one of the first DSL in the history -- comes itself in different shapes. Figure~\ref{fig:sqlshapes} shows three of these shapes on the same basic query example. The top part of the figure shows the plain SQL variant, with a classical \emph{"select, from, where"} clause. In the middle part of the figure, we show the same query written in Java with JOOQ (\url{http://jooq.org}), a fluent API which emphasises its typesafe nature. The lower part of the figure shows again the same query using the Slick API in Scala (\url{http://slick.typesafe.com}).

All shapes of DSLs have strengths and weaknesses whoever you are -- a developer or a user of a DSL. These SQL shapes illustrate this situation. The plain SQL version is an external DSL, making it easier for database experts to write complex queries, but making harder the software engineering job of integrating the DSL with other programming languages. On the entire other side, the JOOQ API is a Java internal DSL. As Java does not provide enough means to host DSL, the best result is a fluent API which mimics the SQL statement in successive method calls. Some of the SQL concepts are clearly recognizable by SQL experts, but some constructs, such as the AND clause, may lead to scoping errors. The job of the DSL developers is also reduced to an API implementation. The third example in Slick shows what can be achieved when the host language, here Scala, has some powerful constructs, such as "filter", that can be reused. The syntax is then less close to the original SQL, but easier for the average Scala developer. Another solution could have been to use the syntactic flexibility of Scala to closely mimic SQL, but this would not have suppressed some drawbacks, i.e., the internal DSL is a leaky abstraction: some arbitrary code may appear at different places in the domain scope~\cite{FowlerDSLBook,mernik2005}, and its  concrete syntax can pose a problem for domain experts or non programmers~\cite{stefik2013}. 

The basic trade-offs between internal and external DSLs have been already identified and are subject to extensive discussions and research for several years. A new trend though is observed. DSLs are now so widespread that they are used by very different users with separate roles and varied objectives. In the case of SQL, users can be marketers, database experts, system administrators, data warehouse managers, software engineers, or web developers. The objectives can vary a lot: prototyping of basic queries to search some data, sophisticated integration of SQL queries into a web application, etc.  
\eemph{Depending on the kinds of users, roles or objectives, an external shape or an internal shape of a language might be a better solution.} Most of the new learners of SQL (e.g., students) have started to learn SQL with explanations and examples written in the external shape of the language as well as a dedicated interactive environment. In the opposite, software engineers have somehow the need to use an internal shape when integrating database concerns.

This diversity poses a major challenge for the DSL engineering discipline: How to provide the good shape of a DSL according to the needs of a user? Our vision for the future of DSL engineering is that the discipline, its foundations, methods, and tools, should go beyond the \emph{polymorphism} which is imposed to the shapes that can take a DSL for its respective various users. 

SQL is not an isolated case. We directly faced similar issues on building and evolving a DSL, called FAMILIAR~\cite{FAMILIAR}.  Similarly to SQL, a diversity of shapes emerged and does make sense for a diversity of users, roles or objectives. We report in the following section our experience in facing such diversity. Grounded in this experience, we claim that developers or users of DSLs should not have to choose their sides: a DSL should be \emph{metamorphic} and change its shape accordingly!

%------------------------------------------------

\section*{A Polymorphic Journey: The FAMILIAR experience}

During four years, we have continuously designed, developed, used, and maintained different solutions for \emph{managing} feature models on a large scale. 
Feature models are by far the most popular notation in industry for variability modeling~\cite{benavides2010}. The formalism can be seen as a technology-independent, high-level, formal representation of options (features) and constraints of a configurable system (e.g., Linux) or a family of products, also called software product lines~\cite{apel2013book}. 

As we will see, the FAMILIAR case study is representative of the "diversity" phenomenon we point out: multiple kinds of users, each with different objectives and needs, may potentially have to manage feature models. The purposes of feature models vary as well. Here is a non-exhaustive list of feature modeling users, together with an usage example: 
\begin{itemize}
\item marketing engineers when characterizing the options (and their constraints) of products offered to customers ; 
\item end users visualising feature models through an intuitive and interactive interface, a (Web) configurator ;
\item system engineers in charge of identifying common and reusable components ;  
\item experts in model checking that use the feature modeling formalism for efficiently verifying complex properties of a configurable system ;  
\item software engineers, for example, web developers in charge of deriving a web configurators from the feature model.
\end{itemize}
The diversity of usage impacts the important properties a solution should exhibit (e.g., learnability, expressiveness, reusability, usability, performance). Intuitively, the emphasis is likely to be more on usability and learnability for non programmers and on ease of integration and performance for software engineers. \eemph{Our story is that we developed "many shapes" of a language to mirror "many forms" of usage, questioning us to the polymorphous nature of our solution.}

The story of FAMILIAR started at the beginning of 2010. At that time, the effort of researchers was mostly centered around the design of a graphical and textual language with a formal semantics for specifying feature models, and the development of \emph{efficient} reasoning operations. On top of the feature modeling languages, numerous Java APIs arose for computing properties of the model.
Naturally we looked at existing Java APIs (SPLAR, FeatureIDE, FAMA, etc.) that provide operations, based on different kinds of solvers (SAT, CSP, BDD). In terms of readability and learnability the APIs were rather complex to apprehend for a programmer (non expert in feature modeling) or a non programmer. The specific audience -- people that want to develop efficient heuristics and have a fine-grained control over the internal details of the solvers -- leads to an API design that did not fit all our requirements.

\begin{figure*}
\center
\includegraphics[scale=0.3]{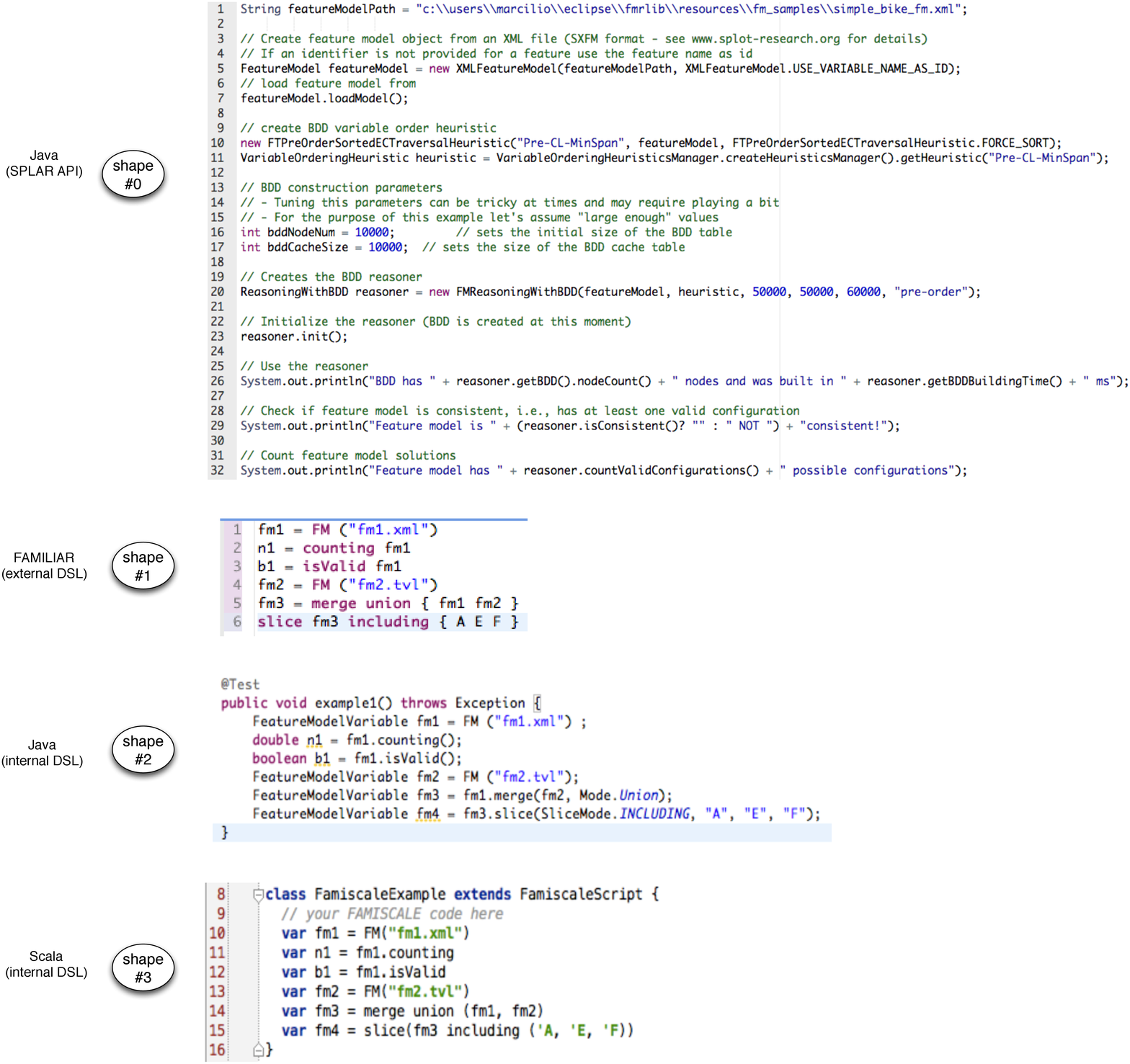}
\caption{An example of SPLAR API and three polymorphic variants of the same domain-specific code snippet}
\label{fig:fmlshapes}
\end{figure*}

We aimed at providing a much more concise solution with less boilerplate, less technical details about the loaders and the solvers, and also more focused on the essential concepts users want to manipulate -- feature models, configurations, and operations on the feature models themselves. 
In the meantime, we aimed at providing novel operations and a comprehensive environment for reverse engineering, composing, decomposing, and managing (in a broad sense) large-scale feature models.

\subsection*{\#1 First try: FAMILIAR, an external DSL}

We thus decided to move to a DSL. We created FAMILIAR~\cite{FAMILIAR} (for FeAture Model scrIpt Language for manIpulation and Automatic Reasoning) an external DSL with its own syntax and its own environment (editors, interactive console, etc.). Figure~\ref{fig:fmlshapes} (page~\pageref{fig:fmlshapes}) gives an example of a FAMILIAR code, see shape \#1. The first three lines almost implement the same behavior than the Java code written in SPLAR (see shape \#0 in the same figure): loading of a feature model, counting of number of configurations and checking consistency properties. All the results are stored into variables (respectively fm1, n1, and b1). We also developed new facilities for supporting numerous formats and composing ("merge") or decomposing ("slice") feature models. Lines 4 to 6 of the FAMILIAR script load a new feature model (fm2) and compute two new models (fm3 and fm4).

The conciseness of the solution, the manipulation of specific abstractions and syntax, the hiding of solver’s details, the absence of boilerplate code, all are arguments in favor of a larger adoption. We report on some experiences hereafter to illustrate why the shift to a DSL appears to be beneficial. In the context of collaborative research projects in different application domains (military applications, video analysis, system engineering, web development) we collaborated and exchanged with experts usually unfamiliar with feature models and/or programming. Using FAMILIAR facilitated the communication during the meetings or brainstormings. Besides we taught to MSc students in Belgium and France an advanced introduction to feature modeling. Again using FAMILIAR facilitated the communication during the courses and lab sessions. The code we presented and discussed was only about feature modeling ; a well-defined vocabulary (e.g., counting, isValid) unifies the terminology and facilitates a common understanding.
Students also experimented advanced aspects of feature modeling like the "merge" and "slice" with a specific environment. In both situations, users can interact with a read-eval-print loop (REPL) to have an immediate evaluation, possibly visual, of the result of the operations. Users can modify of the scripts, if needs be, and repeat the process. REPL and the scripting nature of FAMILIAR allow users to manually verify that the behavior is indeed what is expected, e.g., what have been understood for students.
 
We also used FAMILIAR code snippet into our scientific publications to ease our explanations (the vocabulary of the language is based on the terminology used by scientists)  and to develop other advanced concepts. So far we cannot claim that FAMILIAR was crucial or simply better than existing solutions -- evaluating the properties of a DSL is a very complex activity and a hot research topic \emph{per se} we and others are currently investigating. But clearly, as industrial consultants and collaborators, teachers, or scientists, we identified advantages of using the external DSL and its dedicated environment.      

\subsection*{\#2 The (not so surprising) shift to a fluent Java API}

So everything is done in a brave new world? Unfortunately no. The \emph{external} nature of the DSL posed two kinds of problems.

First, the integration of FAMILIAR into complete applications was trickier than expected. For instance, we used FAMILIAR to model variability of video surveillance processing chains, with the ultimate goal of re-configuring the different vision algorithms according to contextual changes. Calling the FAMILIAR interpreter at runtime -- each time a reconfiguration is needed -- was clearly not a solution due to performance issues. Programming the "glue" between the adaptation logic, as provided by FAMILIAR, and the actual algorithms of the application was also a problem. We also faced integration problems when deploying configurable scientific services into the grid or when generating web configurators. We needed to operate over feature models and configurations, but in practice, the isolated nature of the external DSL leads to difficult or inefficient connections with the outside world and other software ecosystems (e.g., Java). 

Second, the specific boundary of the language naturally limits its expressiveness. It is a desirable property of a DSL to restrict the users to essential language constructs and concepts of a domain. However there may be cons; and it was the case for FAMILIAR. Users wanted to iterate over a collection of feature models, e.g., to rename some features; to check some properties and execute a specific operation, e.g., if the feature model is consistent, the "merging" with another is performed; to reuse some procedures, etc. We added to FAMILIAR some constructs of a general purpose language, like a "foreach"-like loop, an "if-then-else", and ways to define and call reusable scripts. We also added basic facilities for manipulating strings and integers -- something offered for free by any general purpose language. Overall we found no elegant solution for comprehensively giving to users the right expressiveness, i.e., so that the language covers all feature modeling scenarios. And eventually we did not want FAMILIAR to resemble a general purpose language.

Due to the lack of integration and expressiveness of the external DSL in some situations, we developed a solution on top of a general purpose language. Specifically we designed a \emph{fluent} Java API with the goal of being as close as possible to the initial syntax of FAMILIAR. Figure~\ref{fig:fmlshapes} (page~\pageref{fig:fmlshapes}) gives the corresponding Java code of the initial FAMILIAR script (see shape \#2). The API has the merit of being concise and provides idiomatic facilities for loading a feature model and executing reasoning operations. But compared to FAMILIAR scripts, we identified the following drawbacks. First, the distance with the original syntax leads to less concise code and more boilerplate code. Second,  the boundary of the proposed solution is less rigid; domain-specific abstractions are as accessible as internal details of solvers or constructs of a general purpose language. For instance, users run the risk of developing inefficient operations instead of reusing existing ones. It may be relevant for some kinds of users but the experience can be disconcerting for others, e.g., new learners such as students.  Third, no specific environment, e.g., with REPL and graphical editors, has been developed, thus making complex the realization of scenarios in which users interactively play with the models and the operations.        

\subsection*{\#3 Yet another attempt: going internal with FAMISCALE}

To overcome some of the previous limitations, we then aimed at reducing the gap with the original FAMILIAR syntax while preserving some integration capabilities with a general-purpose programming language. We thus developed yet another solution (another internal DSL). We build FAMISCALE on top of the Scala language that notably supports a flexible syntax, implicit type conversions, call-by-name parameters and mixin-class composition. Compared to the Java solution, the gap with FAMILIAR is largely reduced while the Scala interactive interpreter and integration facilities are directly reusable.

The corresponding FAMISCALE code of our feature model example is depicted in the lower part of Figure~\ref{fig:fmlshapes}, page~\pageref{fig:fmlshapes} (see shape \#3). It clearly shows the nearness with the original FAMILIAR syntax, but still, one cannot directly interpret some FAMILIAR code. During some recent application, we faced the need to compute some additional metrics on feature models. Coding and integrating them in the software toolchain was very fast and the newly developed functionalities could be added to a future release seamlessly. On the contrary, using the advanced concepts of the Scala host language was a considerable engineering effort and the whole solution is not that close to FAMILIAR. For example handling variable namespaces and parameterized scripts led to a very different solution from FAMILIAR, but with the benefit of a better integration with the programming side of the Scala platform.

\subsection*{All wrong? No. Metamorphic!}

So what is the best solution? The fluent Java API? The scripting language (FAMILIAR)? The internal DSL in Scala (FAMISCALE)? We considered various tradeoffs, e.g., learnability, expressiveness, reusability, throughout our exploration journey. Our experience is that there is no one-size-fits-all solution or clear superiority of a solution compared to another.  We have a better expressiveness with a Java API, but a more difficult solution to apprehend.  We have a better interactive environment with FAMILIAR but we cannot realize all feature modeling scenarios -- we even doubted at certain points of our development that FAMILIAR \emph{is} a DSL. 

So, are we all wrong from the start?  We may have missed a solution that outperforms all the others; or a future programming technology might emerge to develop such a solution. Yet, our experience is that it does make sense to continue the maintenance of external and internal solutions. Rather than choosing between an internal DSL or an external DSL, we argue that both solutions are eventually relevant.

\begin{sidewaysfigure}
\center
\includegraphics[scale=0.155]{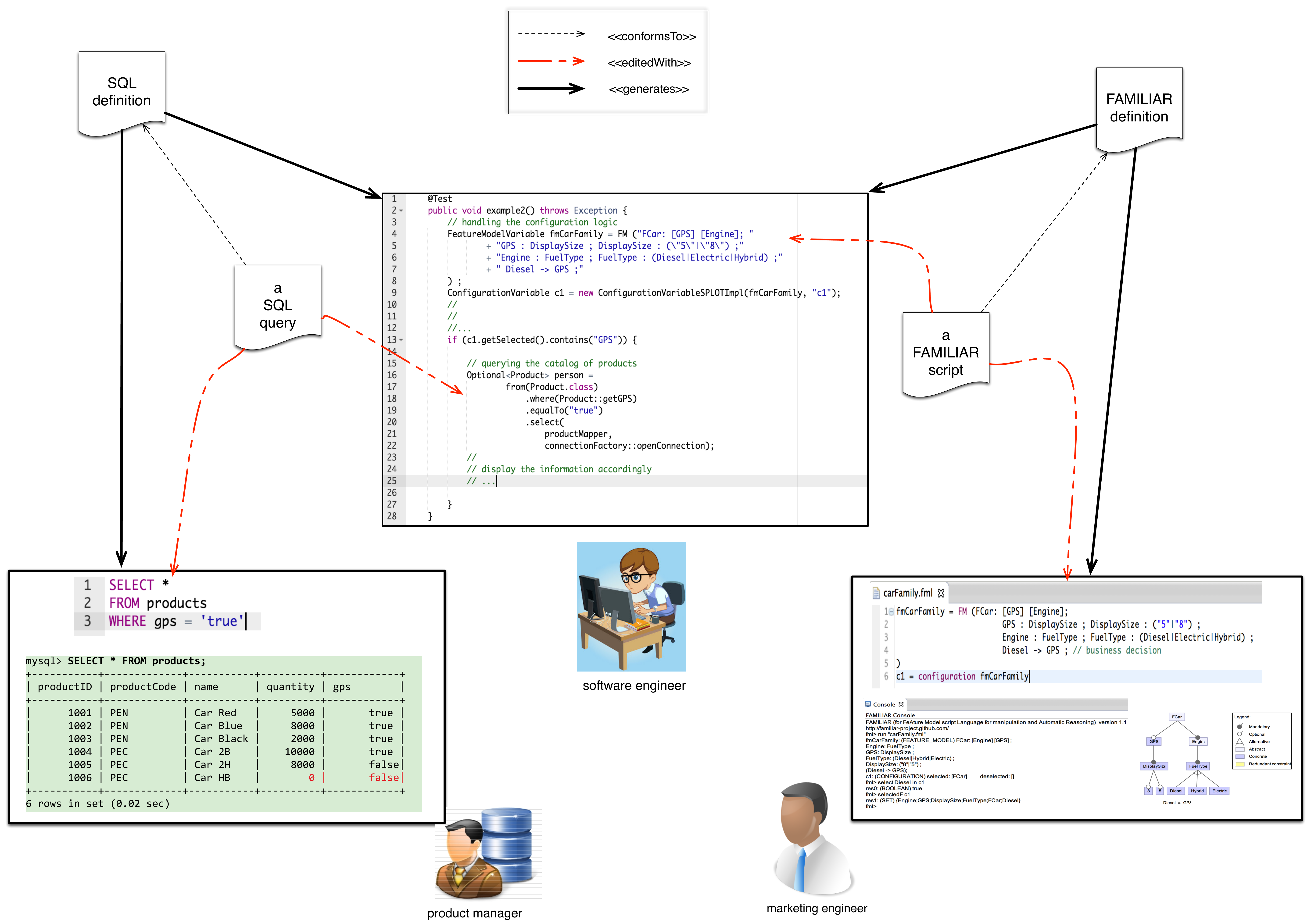}
\caption{Illustration of the vision with a metamorphic SQL, a metamorphic FAMILIAR, and their possible integration\label{fig:metamorph}}
\end{sidewaysfigure}

For instance, new learners of feature models benefit from playing with a specific environment and a dedicated, epurated syntax (FAMILIAR). Software engineers make a far better use of the internal DSL to integrate the feature modeling logic as part of their software project. Why forcing a software engineer to use an external DSL? or a student to start with a Java API? Even for a specific kind of user, the different shapes of the language are likely to be useful. For instance, software engineers themselves can prototype a scripting solution with the external DSL, play with the FAMILIAR environment, and eventually get back to the "internal" shape. Different stakeholders can also be part of the process while using different shapes of the solution (see Figure~\ref{fig:metamorph}). A marketing engineer will more likely use the external DSL to characterize the variability of the system under design, to control some properties of the feature model, etc. Once achieved, the software engineer in charge of  connecting the different feature models will go on with the development.

We argue that this situation and scenarios are generalisable in many DSL engineering contexts, beyond SQL and FAMILIAR. Our key point is that \eemph{the "best" shape of a language heavily depends on the usage context (tasks to perform, kinds of users, etc.): in the FAMILIAR experience, any existing shape we developed has, at some points, some qualities that others do not}.

\section*{Metamorphic DSL: A Vision For the Future of Languages}

Some recent work already started to wipe off the gap between the different shapes. For example, some approaches for implementing external DSLs try to incorporate the benefits from internal DSL by supporting the reuse of already implemented languages and tools. For example Xtext supports the reuse of language libraries such as Xbase, which come with their whole tooling such as editor, type checker and compiler~\cite{efftinge2012}.  Conversely, some approaches try to limit the scope of the host infrastructure that can be reused in an internal DSL to ensure a well-defined isolation of the domain-specific constructs. For example, LMS relies on the staging mechanism to define an external DSL on top of the Scala host infrastructure~\cite{rompf2010}. Besides, projectional language workbenches such as Jetbrain's MPS support the projection in a shape of a purely external DSL or as embedded in a host language similarly to internal DSLs~\cite{voelterMPS}. More generally different strategies for embedding a DSL have been proposed~\cite{tratt2008,erdweg2011,hofer2008}.

These recent efforts attempt to integrate the advantages of the different approaches, progressively wiping off the gap between them. This is an essential experience to master the various possible bridges and differences between the different shapes. Nevertheless, the same concern (variability or SQL queries in our example) usually flows through the life cycle, being addressed by different users with their specific points of view and objectives  (see Figure~\ref{fig:metamorph} for an illustrative scenario). Each user expects to manipulate the same programming artefact through the most appropriate shape of the DSL (incl., the whole tooling). For example, a product manager will manipulate SQL queries with plain text SQL and dedicated tools while a software engineer will use an internal DSL.

Beyond the unification of the different approaches, \eemph{the vision that we foresee for the future of software languages is their ability to be self-adaptable to the most appropriate shape (including the corresponding IDE) according to a particular usage or task. We call \emph{metamorphic DSL} such a domain-specific language, able to change from one shape to another shape}. From the same language description, we envision the ability to derive various IDEs that can be used accordingly. The challenge consists into withstanding the manipulation of the same artifact from the different points of view, each one with their specific representation as well as integration into a host infrastructure.

In our vision, the same programming artifact could be started in isolation so that a stakeholder could describe her concern with a highly dedicated environment. The same artifact would flow (e.g., refinement, transformation, composition, consistency checking) and be combined to the other concerns until eventually obtaining the final system.  The vision we propose does not conflict with the use of multiple languages. It is rather a way to support their \emph{integration} for a coordinated development of diverse domain-specific concerns of a system.

The scenario of Figure~\ref{fig:metamorph} illustrates the vision on two DSLs, namely SQL and FAMILIAR. 
It involves different stakeholders (marketing engineer, product manager, software engineer) that aim at providing a configurator of sales product. Each DSL provides two IDEs from the same language definition (incl., one shared by the two DSLs) that can be indifferently used to edit the same conforming artefact. A shared IDE is used as a common infrastructure to integrate the two concerns (e.g., the integration of FAMILIAR and SQL in Java).

The metamorphic vision raises many challenges. One must still find solutions for the integration of domains, especially between business and technical domains. Systematic methods for evaluating \emph{when} a shape of a DSL meets the expected properties (e.g., learnability) and is more suitable to another would benefit to developers of DSL, but are far from being complete.
The vision also gives rise to questions about the modularization of artifacts. The technical challenge is to share some information, while being able to visualize and manipulate an artifact in a particular representation and in a particular IDE. A global mechanism must ensure consistency of the artifacts between these heterogeneous IDEs.
 
The ability to shape up a DSL would open a new path for an effective communication between humans and the realization of global software engineering scenarios: could Metamorphic DSLs bring language engineering to the next level, enabling user-driven task-specific support in domain-specific worlds?

\subsection*{Acknowledgments} We thank our colleagues Thomas Degueule, Guillaume B\'{e}can, Olivier Barais, Julien Richard-Foy, and Jean-Marc J\'{e}z\'{e}quel for fruitful comments and discussions on earlier drafts of this paper. 
This work is part of the GEMOC Initiative, and partially funded by the ANR INS Project GEMOC (ANR-12-INSE-0011).

%----------------------------------------------------------------------------------------
%	BIBLIOGRAPHY
%----------------------------------------------------------------------------------------

\bibliographystyle{unsrt}

\bibliography{metamorph}

\begin{thebibliography}{10}

\bibitem{mernik2005}
Marjan Mernik, Jan Heering, and Anthony~M. Sloane.
\newblock When and how to develop domain-specific languages.
\newblock {\em ACM Comput. Surv.}, 37(4):316--344, 2005.

\bibitem{FowlerDSLBook}
Martin Fowler.
\newblock {\em Domain Specific Languages}.
\newblock Addison-Wesley Professional, 2010.

\bibitem{voelter2013}
Markus Voelter, Sebastian Benz, Christian Dietrich, Birgit Engelmann, Mats
  Helander, Lennart C.~L. Kats, Eelco Visser, and Guido Wachsmuth.
\newblock {\em DSL Engineering - Designing, Implementing and Using
  Domain-Specific Languages}.
\newblock dslbook.org, 2013.

\bibitem{stefik2013}
Andreas Stefik and Susanna Siebert.
\newblock An empirical investigation into programming language syntax.
\newblock {\em Trans. Comput. Educ.}, 13(4):19:1--19:40, November 2013.

\bibitem{FAMILIAR}
Mathieu Acher, Philippe Collet, Philippe Lahire, and Robert~B. France.
\newblock Familiar: A domain-specific language for large scale management of
  feature models.
\newblock {\em Science of Computer Programming (SCP)}, 78(6):657--681, 2013.

\bibitem{benavides2010}
David Benavides, Sergio Segura, and Antonio~Ruiz Cort{\'e}s.
\newblock Automated analysis of feature models 20 years later: A literature
  review.
\newblock {\em Inf. Syst.}, 35(6):615--636, 2010.

\bibitem{apel2013book}
Sven Apel, Don Batory, Christian K{\"a}stner, and Gunter Saake.
\newblock {\em Feature-Oriented Software Product Lines: Concepts and
  Implementation}.
\newblock Springer-Verlag, 2013.

\bibitem{efftinge2012}
Sven Efftinge, Moritz Eysholdt, Jan K\"{o}hnlein, Sebastian Zarnekow, Robert
  von Massow, Wilhelm Hasselbring, and Michael Hanus.
\newblock Xbase: Implementing domain-specific languages for java.
\newblock In {\em Proceedings of the 11th International Conference on
  Generative Programming and Component Engineering}, GPCE '12, pages 112--121,
  New York, NY, USA, 2012. ACM.

\bibitem{rompf2010}
Tiark Rompf and Martin Odersky.
\newblock Lightweight modular staging: A pragmatic approach to runtime code
  generation and compiled dsls.
\newblock {\em SIGPLAN Not.}, 46(2):127--136, October 2010.

\bibitem{voelterMPS}
Markus Voelter, Daniel Ratiu, Bernd Kolb, and Bernhard Sch{\"a}tz.
\newblock mbeddr: instantiating a language workbench in the embedded software
  domain.
\newblock {\em Autom. Softw. Eng.}, 20(3):339--390, 2013.

\bibitem{tratt2008}
Laurence Tratt.
\newblock Domain specific language implementation via compile-time
  meta-programming.
\newblock {\em ACM Trans. Program. Lang. Syst.}, 30(6):31:1--31:40, October
  2008.

\bibitem{erdweg2011}
Sebastian Erdweg, Tillmann Rendel, Christian K\"{a}stner, and Klaus Ostermann.
\newblock Sugarj: Library-based syntactic language extensibility.
\newblock In {\em Proceedings of the 2011 ACM International Conference on
  Object Oriented Programming Systems Languages and Applications}, OOPSLA '11,
  pages 391--406, New York, NY, USA, 2011. ACM.

\bibitem{hofer2008}
Christian Hofer, Klaus Ostermann, Tillmann Rendel, and Adriaan Moors.
\newblock Polymorphic embedding of dsls.
\newblock In {\em Proceedings of the 7th International Conference on Generative
  Programming and Component Engineering}, GPCE '08, pages 137--148, New York,
  NY, USA, 2008. ACM.

\end{thebibliography}

%----------------------------------------------------------------------------------------

\end{document}